\documentclass[preprint]{vgtc}               % preprint
\ifpdf%                                % if we use pdflatex
  \pdfoutput=1\relax                   % create PDFs from pdfLaTeX
  \usepackage{graphicx}                % allow us to embed graphics files
  \DeclareGraphicsExtensions{.pdf,.png,.jpg,.jpeg} % for pdflatex we expect .pdf, .png, or .jpg files
\else%                                 % else we use pure latex
  \usepackage{graphicx}                % allow us to embed graphics files
  \DeclareGraphicsExtensions{.eps}     % for pure latex we expect eps files
\fi%

\graphicspath{{figures/}{pictures/}{images/}{./}} % where to search for the images

\usepackage{microtype}                 % use micro-typography (slightly more compact, better to read)
\PassOptionsToPackage{warn}{textcomp}  % to address font issues with \textrightarrow
\usepackage{textcomp}                  % use better special symbols
\usepackage{mathptmx}                  % use matching math font
\usepackage{times}                     % we use Times as the main font
\usepackage{cite}                      % needed to automatically sort the references
\usepackage{tabu}                      % only used for the table example
\usepackage{booktabs}                  % only used for the table example
\usepackage[utf8]{inputenc}

\usepackage{amsmath}
\usepackage{amsfonts}
\usepackage{amssymb}
\usepackage{dsfont}
\usepackage{mathtools}
\usepackage{subfig}
\usepackage[per-mode=symbol]{siunitx}
\usepackage{xcolor}
\usepackage{tikz}
\usetikzlibrary{positioning, calc}
\usepackage{enumitem}

\usepackage{comment}

\DeclareRobustCommand\circled[1]{\tikz[baseline=(char.base)]{
            \node[shape=circle,draw,inner sep=0.5pt] (char) {\small{#1}};}}

\newcommand{\CHANGE}[1]{\textcolor{orange}{#1}}
\renewcommand{\CHANGE}[1]{#1}

\newcommand{\etal}{et al.}

\usepackage[ruled, vlined, linesnumbered]{algorithm2e}
\usepackage{etoolbox}
\makeatletter
\patchcmd{\@algocf@start}% <cmd>
  {-1.5em}% <search>
  {0pt}% <replace>
  {}{}% <success><failure>
\makeatother

\newlength\mylen

\SetKw{KwArray}{Array}

\DeclareMathOperator{\sad}{sad}

\newcommand{\ptype}{p_{\text{type}}}

\newcommand{\ctype}{c_{\text{type}}}

\DeclareMathOperator{\Bin}{Bin}

\DeclareMathOperator{\Beta}{Beta}

\renewcommand{\vec}[1]{\mathbf{#1}}

\newcommand{\myLabelA}{A}
\newcommand{\myLabelB}{B}
\newcommand{\myLabelC}{C}

\onlineid{0}

\vgtccategory{Technique}

\vgtcinsertpkg

\preprinttext{To appear in an IEEE VGTC sponsored conference.}

\title{Visualizing Confidence Intervals for Critical Point Probabilities\\in 2D Scalar Field Ensembles}

\author{Dominik Vietinghoff\thanks{e-mail: vietinghoff@informatik.uni-leipzig.de}\\ %
    \scriptsize Leipzig University %
\and Michael B\"ottinger\thanks{e-mail: boettinger@dkrz.de}\\ %
    \scriptsize Deutsches Klimarechenzentrum 
\and Gerik Scheuermann\thanks{e-mail: scheuermann@informatik.uni-leipzig.de}\\ %
    \scriptsize Leipzig University %
\and Christian Heine\thanks{e-mail: heine@informatik.uni-leipzig.de}\\ %
    \scriptsize Leipzig University %
}

\teaser{
  \centering
  \begin{tikzpicture}
      \node[draw, black, thick, inner sep=0pt] (leftpic) {%
        \includegraphics[trim={600px 420px 260px 260px},clip,scale=0.14]{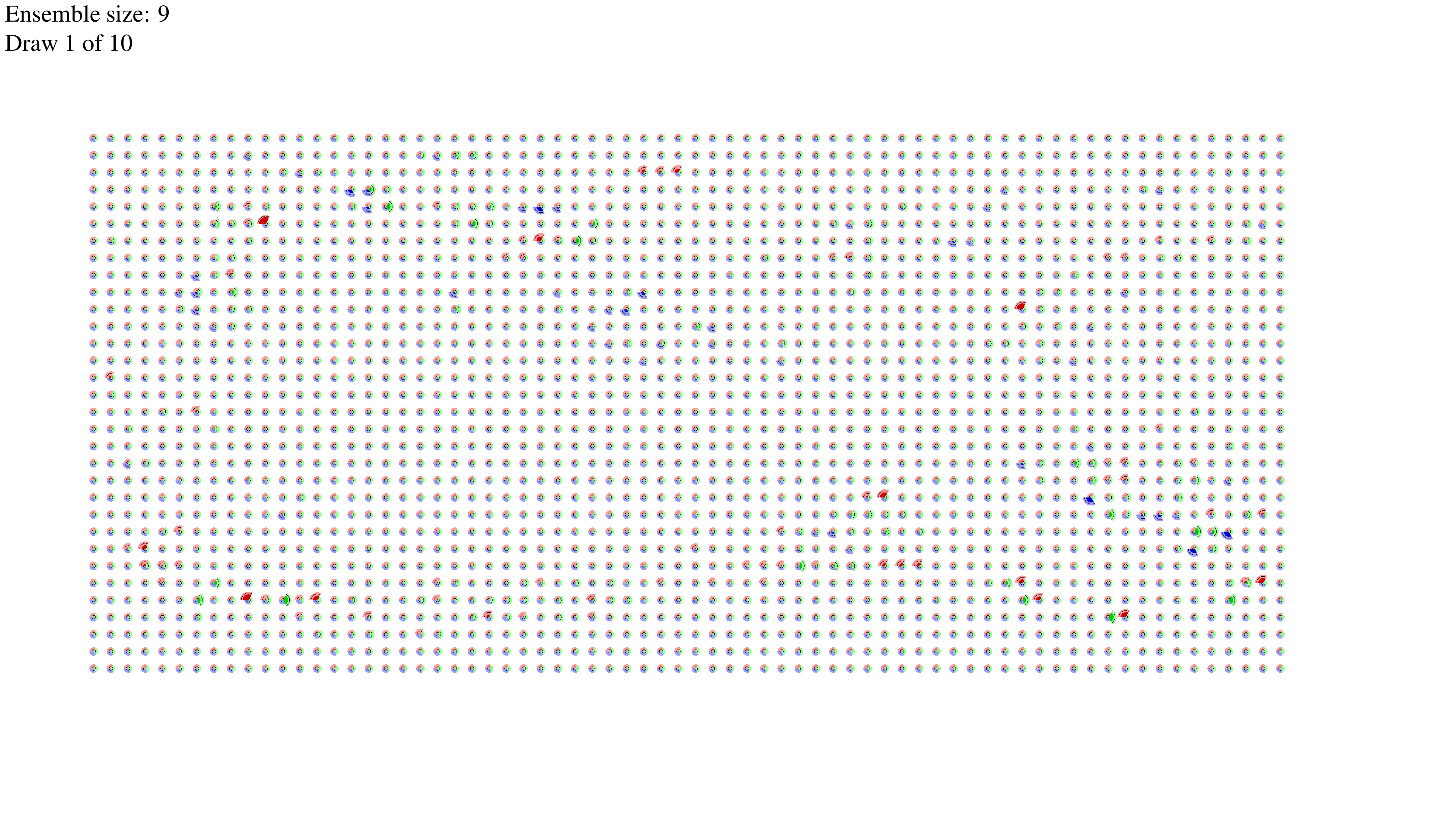}};
      \node[draw, black, thick, inner sep=0pt, right=0.2em of leftpic] (rightpic) {%
        \includegraphics[trim={600px 420px 260px 260px},clip,scale=0.14]{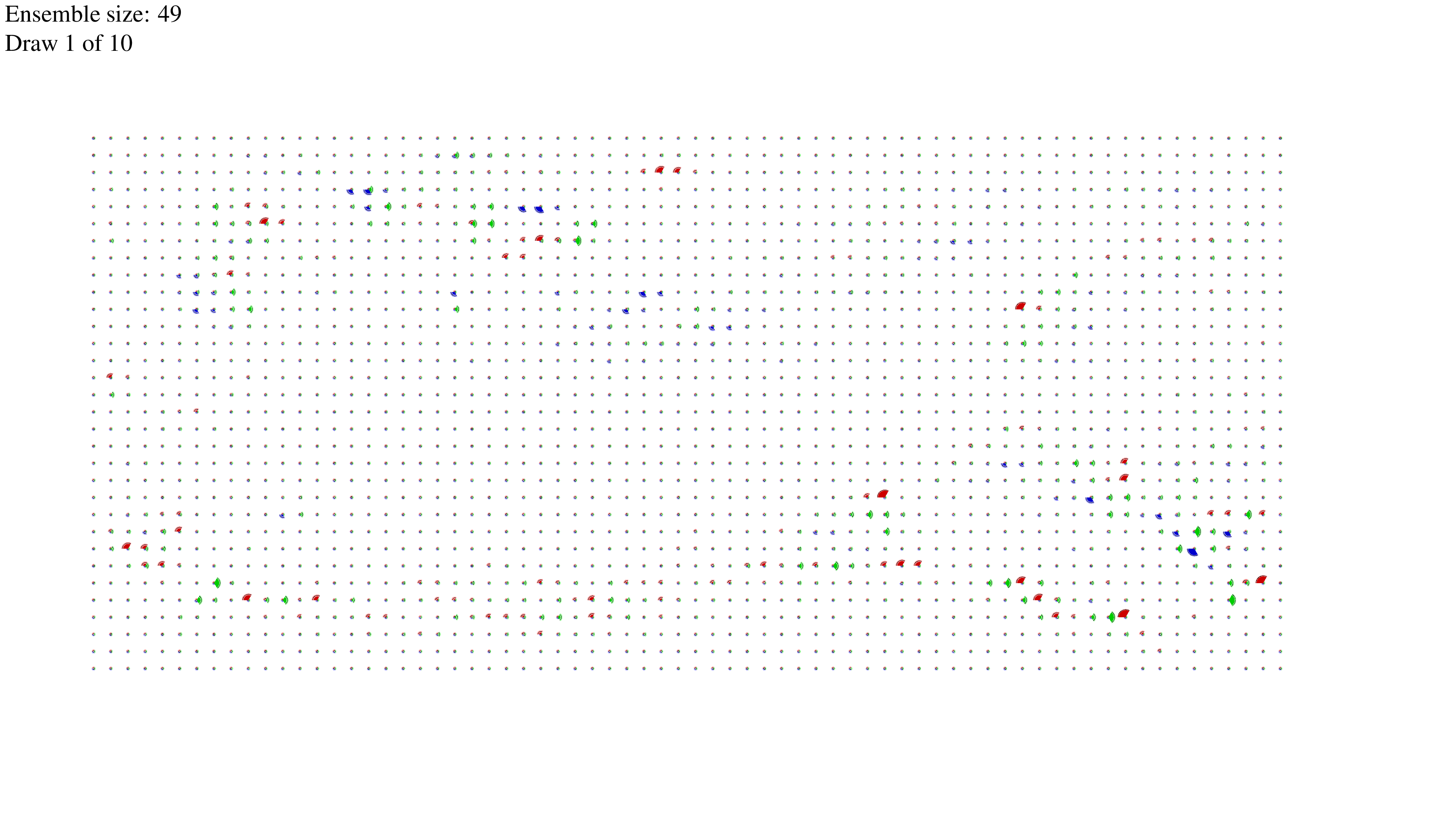}};
      \node[inner sep=0pt, anchor=north] (legend) at ($0.5*(leftpic.south east)+0.5*(rightpic.south west)-(0,0.2em)$) {%
        \includegraphics[scale=0.75]{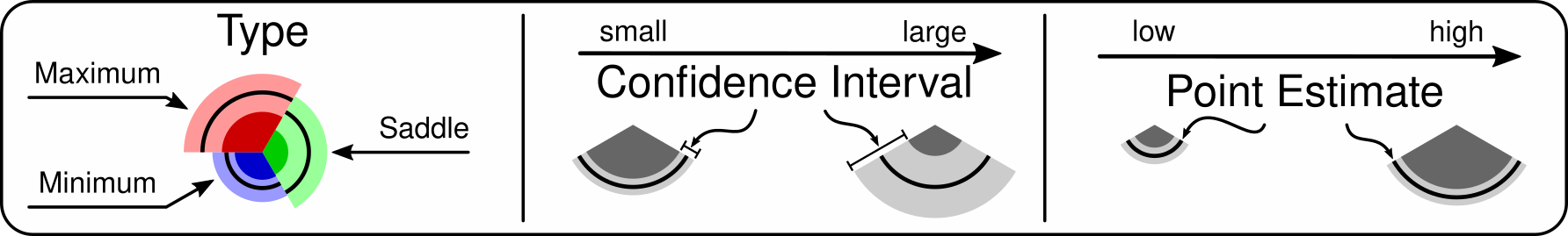}};
        
      \node[draw, black, rounded corners=2pt, minimum width=1.6cm, minimum height=3.3cm, right=0.2em of rightpic.north east, anchor=north west] (legend_outline)  {};
      \node[right=0.05em of legend_outline.west, anchor=north, align=center, text width=3.9cm, rotate=90] (legend_header) {\textbf{Glyph Size (Inset)}};
     
      \node[below right=0.9em and 0.8em of legend_outline.north, anchor=north] (legend25label) {25\%};
      \node[inner sep=0pt, above=0em of legend25label.north, anchor=south] (legend25) {%
         \includegraphics[angle=-90, trim={1490px 640px 806px 630px},clip,scale=0.5]{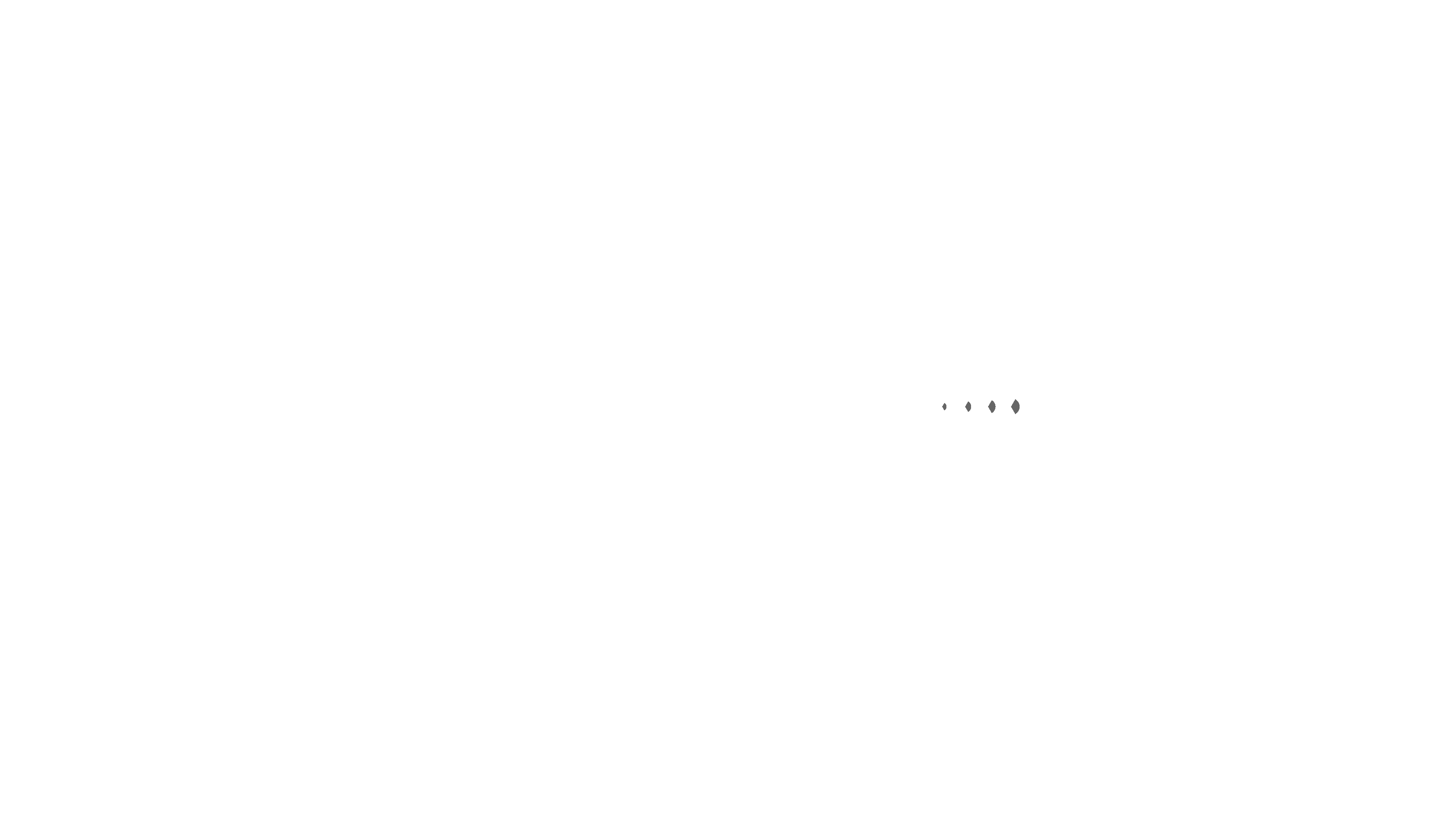}}; 

      \node[below=0.8em of legend25label.south, anchor=north] (legend50label) {50\%};
      \node[inner sep=0pt, below=0em of legend50label.north, anchor=south] (legend50) {%
         \includegraphics[angle=-90,trim={1526px 640px 766px 630px},clip,scale=0.5]{pictures/Results/Synthetic/Synthetic_Legend.png}};
        
      \node[below=0.8em of legend50label.south, anchor=north] (legend75label) {75\%};
      \node[inner sep=0pt, below=0em of legend75label.north, anchor=south] (legend75) {%
         \includegraphics[angle=-90, trim={1563px 640px 728px 630px},clip,scale=0.5]{pictures/Results/Synthetic/Synthetic_Legend.png}};
        
      \node[below=0.8em of legend75label.south, anchor=north] (legend100label) {100\%};
      \node[inner sep=0pt, below=0em of legend100label.north, anchor=south] (legend100) {%
         \includegraphics[angle=-90, trim={1599px 640px 690px 630px},clip,scale=0.5]{pictures/Results/Synthetic/Synthetic_Legend.png}};

      \node[inner sep=0pt, draw, black, very thick, below left=0.2cm and 0.2cm of leftpic.north east, anchor=north east] (leftpicinsert) {%
        \includegraphics[trim={700px 870px 1350px 310px},clip,scale=0.5]{pictures/Results/Synthetic/Jeffreys_Intervals_nSamples009_draw000.png}};
      \node[inner sep=0pt, draw, black, very thick, below left=0.2cm and 0.2cm of rightpic.north east, anchor=north east] (rightpicinsert) {%
        \includegraphics[trim={700px 870px 1350px 310px},clip,scale=0.5]{pictures/Results/Synthetic/Jeffreys_Intervals_nSamples049_draw000.png}};
      
      \node[draw, black, line width=.2mm, minimum width=1.21cm, minimum height=.54cm, inner sep=0pt, below left=0.27cm and 5.4cm of leftpic.north east, anchor=north east] (leftpicregion) {};
      \node[draw, black, line width=.2mm, minimum width=1.21cm, minimum height=.54cm, inner sep=0pt, below left=0.27cm and 5.4cm of rightpic.north east, anchor=north east] (rightpicregion) {};
      
      \node[draw, black, thick, fill=white, rounded corners=2pt, above right=0.2cm and 0.2cm of leftpic.south west, anchor=south west] (leftpiclabel) {9 Samples};
      \node[draw, black, thick, fill=white, rounded corners=2pt, above right=0.2cm and 0.2cm of rightpic.south west, anchor=south west] (rightpiclabel) {49 Samples};
      
      \draw[black,line width=.2mm] (leftpicregion.north east) -- (leftpicinsert.north west);
      \draw[black,line width=.2mm] (leftpicregion.south east) -- (leftpicinsert.south west);
      
      \draw[black,line width=.2mm] (rightpicregion.north east) -- (rightpicinsert.north west);
      \draw[black,line width=.2mm] (rightpicregion.south east) -- (rightpicinsert.south west);

      \node[draw, circle, line width=.12em, fill=white, inner sep=1.2pt, below left=0.25cm and 1.0cm of leftpicinsert.south] (labelA) {\myLabelA};
      \node[circle, minimum width=0.2cm, minimum height=0.2cm] (labelAend) at ($(leftpicinsert.center) + (-1.42cm, -0.7cm)$) {};
      \draw[->, black, line width=.15em] (labelA) to [bend right=15] (labelAend);

      \node[draw, circle, line width=.12em, fill=white, inner sep=1.2pt, below right=0.25cm and 0.05cm of leftpicinsert.south] (labelB) {\myLabelB};
      \node[circle, minimum width=0.2cm, minimum height=0.2cm] (labelBend) at ($(leftpicinsert.center) + (0.55cm, -0.25cm)$) {};
      \draw[->, black, line width=.15em] (labelB) to [bend left=15] (labelBend);
      
      \node[draw, circle, line width=.12em, fill=white, inner sep=1.2pt, below right=0.25cm and 0.0cm of rightpicinsert.south] (labelC) {\myLabelC};
      \node[circle, minimum width=0.55cm, minimum height=0.55cm] (labelCend) at ($(rightpicinsert.center) + (0.55cm, 0.8cm)$) {};
      \draw[->, black, line width=.15em] (labelC) to [bend left=8] (labelCend);
    \end{tikzpicture}
  \caption{Confidence intervals for critical point occurrence probabilities for two ensembles of synthetic fields with different number of members.
  The red, blue, and green thirds of the glyphs show the confidence intervals for maxima, minima, and saddles at the respective grid points. The area of the darker, inner arcs is proportional to the lower bounds of the found confidence intervals, the area of the lighter, outer arcs to their upper bounds. The black line marks the point estimate for the probability of a critical point of each type computed from the input ensemble. Glyphs showing a large portion of a dark color (e.g., \circled{B} and \circled{C}) mark points where a critical point of the respective type can be expected with high confidence (95\% confidence level), whereas glyphs with high uncertainty (due to small sample approximations) feature wider outer arcs in lighter colors (e.g, \circled{A} and \circled{B}) (cf. legend at the bottom).
  }
  \label{fig:teaser}
}

\abstract{An important task in visualization is the extraction and highlighting of dominant features in data to support users in their analysis process. Topological methods are a well-known means of identifying such features in deterministic fields. However, many real-world phenomena studied today are the result of a chaotic system that cannot be fully described by a single simulation. 
Instead, the variability of such systems is usually captured with ensemble simulations that produce a variety of possible outcomes of the simulated process.
The topological analysis of such ensemble data sets and uncertain data, in general, is less well studied. 
In this work, we present an approach for the computation and visual representation of confidence intervals for the occurrence probabilities of critical points in ensemble data sets. We demonstrate the added value of our approach over existing methods for critical point prediction in uncertain data on a synthetic data set and show its applicability to a data set from climate research.} % end of abstract

\CCScatlist{
  Uncertainty visualization, scalar topology, critical points, ensemble data, climate data, inferential statistics, glyphs.
}

\begin{document}

\setlength{\abovedisplayskip}{4pt}
\setlength{\belowdisplayskip}{4pt}

\setlength{\abovecaptionskip}{5pt}
\setlength{\belowcaptionskip}{-7pt}


\firstsection{Introduction}

\maketitle

\begin{figure}[t]
  \centering
  \vspace*{-1em}
  \subfloat[Neighbors]{\includegraphics[width=0.22\linewidth]{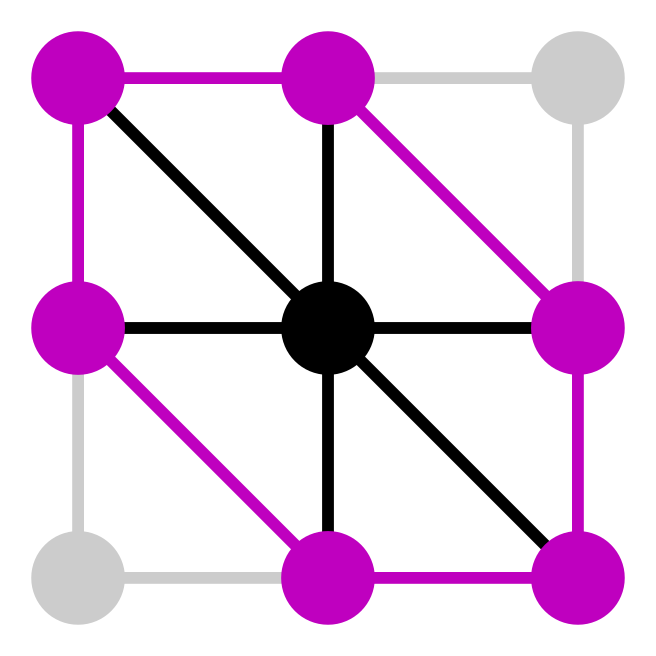}}\quad
  \subfloat[Minimum]{\includegraphics[width=0.22\linewidth]{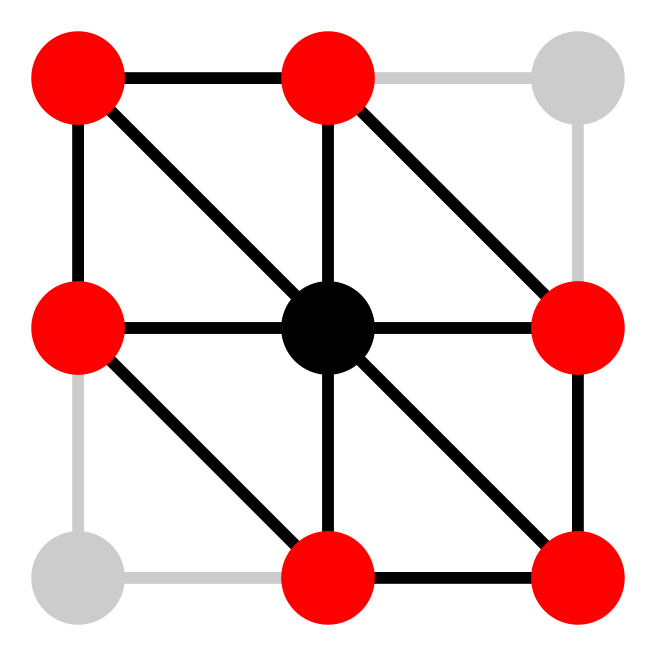}}\quad
  \subfloat[Maximum]{\includegraphics[width=0.22\linewidth]{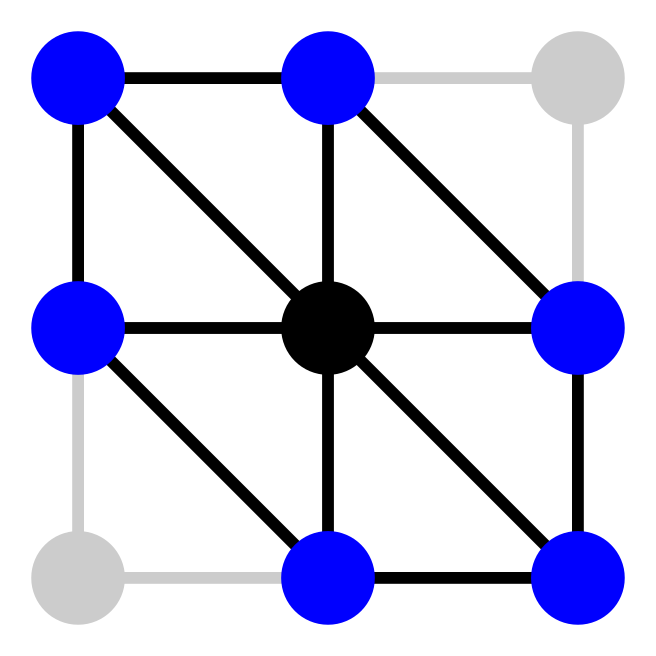}}\quad
  \subfloat[Saddle]{\includegraphics[width=0.22\linewidth]{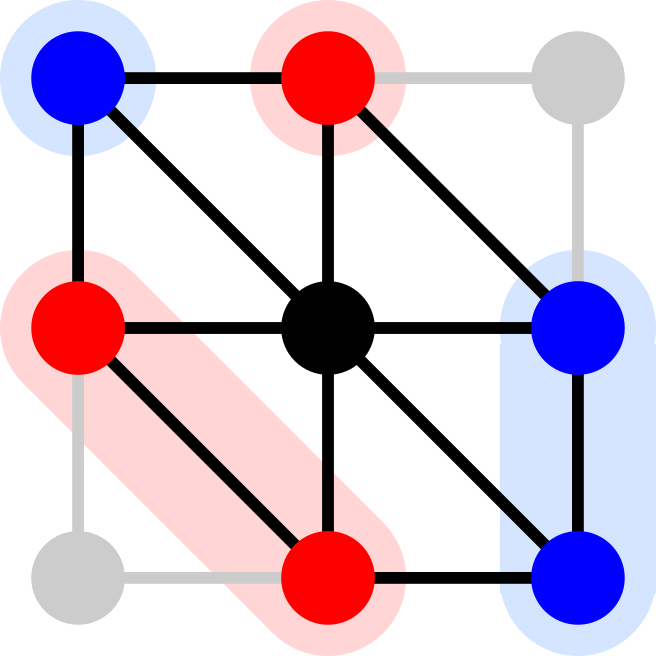}}
  \caption{Critical points are classified based on the values at the direct neighbors of a grid point (purple parts in (a)). Higher and lower function values are represented as red and blue dots, respectively. The grid point is a minimum/maximum if all neighboring points have a larger/lower function value ((b) and (c)). If the neighborhood decomposes into more than two connected components consisting only of points with higher or lower function values (light red and light blue), the point is a saddle point (d).}
  \label{fig:link-critical-points}
\end{figure}

% Captions of first chapter must be defined in main document before `maketitle`
% \section{Introduction}
\label{sec:introduction}
Topological methods are routinely used today to visualize and study specific features in deterministic (scalar) fields. 
An important class of topological features is critical points, as they mark extremal values in the data and indicate points where the level set topology of a field undergoes fundamental changes. In piecewise-linear (PL) scalar fields given on simplicial grids, critical points can only occur at the grid's vertices, and their type is determined solely by the values of their direct neighbors in the grid (cf. Figure~\ref{fig:link-critical-points})~\cite{Tierny2017}.

It is less well understood how topological methods, and in particular critical point detection, can be extended to analyze uncertain data~\cite{Heine2016}.
The related uncertainty can have different sources~\cite{Potter2012}. 
Most existing critical point extraction methods focus on the investigation of \emph{aleatoric uncertainty} by assuming or requiring a latent distribution underlying the data but do not consider \emph{epistemic uncertainties} arising from approximating those distributions by a finite number of input fields (cf. Section~\ref{sec:relatedwork}). 

In this work, we present a method for computing and visualizing confidence intervals for the occurrence probabilities of critical points in ensembles of PL 2D scalar fields. For this, no assumptions on the underlying distribution are made. We evaluate our method with synthetic data (which we also used to compare existing critical point extraction methods in our previous work~\cite{Vietinghoff2022}) and apply it to a real-world data set from climate research.

\section{Related Work}
\label{sec:relatedwork}
The consideration of uncertainty has been identified as one of the most important research directions in visualization~\cite{Johnson2004}. Key challenges in this area were listed by Johnson and Sanderson~\cite{Johnson2003}. In their survey on topological methods in visualization, Heine \etal~\cite{Heine2016} also found a lack of methods for the treatment of uncertain data.

The progress in the field of uncertainty visualization has been summarized in multiple state-of-the-art reports~\cite{Bonneau2014, Brodlie2012, Potter2012, Wang2019, Kamal2021}. Here, we focus on works closely related to our own in that they also investigate critical points in uncertain scalar fields.
We group them into two categories based on the type of considered uncertainty following the classification used by Potter \etal\cite{Potter2012}:

\textbf{Aleatoric uncertainty.}
This kind of uncertainty is inherent to the (physical) process under investigation. The exact state of a chaotic system at some point in time, for instance, is difficult to predict as small perturbations to the initial state cause a different evolution over time. This behavior can often be modeled by a random variable following some probability distribution. Different methods were proposed to capture this type of uncertainty during critical point detection:
Mihai and Westermann~\cite{Mihai2014} compute confidence intervals for the gradient as well as for the trace and the determinant of the Hessian matrix to find points that are likely to be critical points of some type. This, implicitly, is done by testing for the hypothesis whether the gradient is the zero vector (indicating a critical point) or not. In their work, Günther \etal~\cite{Guenther2014} investigate stochastic processes whose probability density has finite support. They find \emph{mandatory critical points}, that is, regions in the domain where every realization of the stochastic process must have at least one critical point of a certain type. Liebmann and Scheuermann~\cite{Liebmann2016} introduce the \emph{realization space}, a high-dimensional space in which each point uniquely describes a realization of the stochastic process. In this space, they identify regions of constant neighbor configuration, which they call \emph{patches}. For singular patches, i.e., patches with a neighbor configuration corresponding to a critical point, they then compute the accumulated probability from all realizations contained in the patch and propose a tracking mechanism that allows merging neighboring singular patches of consistent type to larger regions.

\textbf{Epistemic uncertainty.}
This kind of uncertainty arises from a lack of knowledge or limited data, e.g., if an ensemble consists of an insufficient number of members to fully capture the properties of the underlying stochastic process. That is, epistemic uncertainties often additionally arise while trying to capture aleatoric uncertainties by means of ensemble simulations. The following methods---while also investigating the aleatoric uncertainty of critical points as the methods discussed in the previous paragraph---also take into account epistemic uncertainties due to finite sample approximations.

Using a different test statistic, Vietinghoff \etal~\cite{Vietinghoff2021} extended the approach by Mihai and Westermann~\cite{Mihai2014} to find likely positions of critical points in the (unknown) expectation value field of the underlying stochastic process.
In recent work, Vietinghoff \etal~\cite{Vietinghoff2022} used Bayesian inference to estimate the probability density for the occurrence probabilities of critical points.
They used the expected value of that density as a point estimator for the occurrence probabilities of critical points and computed its differential entropy as a measure of the reliability of that estimator. A comparison with the previously mentioned methods on synthetic data showed that this approach reduced the risk of the viewer to rely on results that are merely a product of chance.
This underlined the importance of the consideration of variability due to small sample approximations.
In this work, we aim at an even more informative representation of such uncertainties by means of confidence intervals for those probabilities.  

\section{Method}
\label{sec:methods}
\subsection{Basic Concepts}
\label{seq:concepts}
Consider a random variable (RV) $X$ drawn from a distribution $F$.
A common task in statistics is to find certain properties $\theta$ of $X$. The exact distribution of $X$ is often unknown; only a finite number of (independent) observations $X_{1},\dots, X_{m}\sim F$ is available.
In such cases, one therefore approximates $\theta$ with some \emph{estimate} $\hat{\theta}$ computed from those observations. 
As $X_1,\dots, X_m$ constitutes only a finite sample from $X$'s distribution, $\hat{\theta}$ will deviate more or less from the true value of $\theta$ depending on the number of considered observations~$m$. 

The variability of $\hat{\theta}$ makes it risky to put trust in this \emph{single point estimate}.
In addition to this value, one, therefore, often specifies a \emph{confidence interval} (CI) that will hold the true value of $\theta$ with a certain probability. More formally, let $\gamma \in [0,1]$. 
We then call $[a,b]$ a CI with confidence level $\gamma$ for $\theta$ if the probability of $[a,b]$ to contain $\theta$ is at least $\gamma$: $\mathds{P}(\theta\in [a,b]) \geq \gamma$. Note that in this, $\theta$ is a fixed (but unknown) parameter, and it is rather the interval $[a,b]$ that varies with repeated samples, thus containing $\theta$ in only $100\%\cdot \gamma$ of such repetitions. More details on the general concept of CIs can be found, e.g., in the book by Wasserman~\cite[Section 6.3.2]{Wasserman2013}.

\subsection{CIs for Occurrence Probabilities of Critical Points}
\label{sec:math-problem-descr}
We are interested in the occurrence probabilities of critical points in ensembles of PL scalar fields $\vec{f}_1, \dots, \vec{f}_m$ defined on a common 2D simplicial grid, where each $\vec{f}_i$ is the vector of the field's values at the grid vertices. For each field $\vec{f}_i$, a grid point $\vec{x}$ is either of some $\mathrm{type} \in \{\max, \min, \sad\}$ (cf. Figure~\ref{fig:link-critical-points}) or not, e.g., $\vec{x}$ is either a maximum or not. 
\CHANGE{The occurrence of those events can hence be described as $m$ independent Bernoulli trials taking a value of 1 (success) if $\vec{x}$ is of the respective type and 0 (failure) otherwise. The number of occurrences of each type at $\vec{x}$ in an ensemble of size $m$ can then be expressed as binomial RVs $X_{\text{type}} \sim \Bin(m, \ptype)$, where $\ptype$ is the (unknown) success probability of the Bernoulli experiments, i.e., the probability for $\vec{x}$ to be of that type. Note that while we assume the fields' values at a single grid point to be stochastically independent, we do not require the RVs $X_{\text{type}}$ to be independent across different types or grid points, which they generally are not. Our goal is to reason about the aleatoric uncertainty of critical points by estimating the occurrence probabilities $\ptype$. A common estimate for the success probability of a binomial distribution is the relative frequency
\begin{equation}
    \label{eq:point-estimate}
    \hat{p}_\text{type} := \frac{\ctype}{m},
\end{equation}
where $\ctype$ is the number of fields in which $\vec{x}$ has that type~\cite{Wasserman2013}.}
\CHANGE{Estimating $\ptype$ from observations produces epistemic uncertainties due to finite-sample errors that can be quantified with CIs.}

Different methods for the approximation of CIs for the success probability of binomial distributions exist~\cite{Brown2001}. Here, we use \emph{Jeffreys intervals}, as they were found to be well suited for small sample sizes of $m \leq 40$ but also produce results comparable to those of other methods for larger samples. 
This approach uses Bayesian inference~\cite[Chapter~11]{Wasserman2013} and the Jeffreys prior~\cite{Jeffreys1961} to find that a likely distribution of $\ptype$, given $\ctype$, is the beta distribution $\Beta(1/2 + \ctype, 1/2 + m - \ctype)$.
Then 
\begin{align}
\label{eq:confidence-interval}
\begin{split}
        [\ptype^{\text{lower}}, \ptype^{\text{upper}}]:= [ &B( \alpha/2; 1/2 + \ctype, 1/2 + m - \ctype), \\
     &B(1 - \alpha/2; 1/2 + \ctype, 1/2 + m - \ctype)],
\end{split}
\end{align}
is an equitailed CI (i.e., $\mathds{P}(\ptype < \ptype^{\text{lower}}) = \mathds{P}(\ptype > \ptype^{\text{upper}})$) for $\ptype$. In this, $\alpha := 1-\gamma$ is the probability that the interval does \emph{not} include the true value of $\ptype$ and $B(\,\cdot\,; a_1, a_2)$ denotes the quantile function of the $\Beta(a_1, a_2)$ distribution. The equitailedness is another positive property of the Jeffreys interval which is not generally given for other CIs.
To avoid ill-behaved CIs in the case that $\ptype$ is close to zero or one, we follow the suggestion by Brown \etal~\cite{Brown2001} and set $\ptype^{\text{lower}} = 0$ if $\ctype = 0$ and $\ptype^{\text{upper}} = 1$ if $\ctype = m$.

\subsection{Visual Design}
\label{sec:visual-design-dominik}
The above procedure results in nine values (point estimate, Equation~\ref{eq:point-estimate}, upper and lower CI boundary, Equation~\ref{eq:confidence-interval}, for each type) for each grid point. To be able to display all those values at each grid location, we propose the following approach: At each grid point $\vec{x}$, we draw a sunburst-like glyph as illustrated by the legend at the bottom of Figure~\ref{fig:teaser}. The glyph consists of three regions, one for each critical point type. 
For each type, we plot the point estimate as black arc with enclosing area proportional to $\hat{p}_\text{type}(\vec{x})$ (thus the radius is proportional to $\sqrt{\hat{p}_\text{type}(\vec{x})}$). 
We further visualize the upper bounds of the CIs as circle segments with areas proportional to $\ptype^{\text{upper}}(\vec{x})$ filled with a light shade of red, blue, or green for maxima, minima, and saddle points, respectively. On top of those segments, we then plot the lower bounds of the CIs as circle segments with areas proportional to $\ptype^{\text{lower}}(\vec{x})$ in darker shades of those colors. 
The glyph marked \circled{\myLabelC} in Figure~\ref{fig:teaser} shows a large, dark blue segment and a narrow, light blue arc that can be interpreted as a likely minimum with low uncertainty. 
Glyph \circled{\myLabelB}, on the other hand, has the highest probability of being a maximum (black arc and dark red circle segment) but also shows a wide, light red arc indicating that no precise statement on the true maximum probability can be made for that grid point.

This glyph-based approach allows for a first intuition on the likely positions of critical points and serves as an indicator for the uncertainty due to small sample sizes.
As it is not possible to read absolute probability values directly from the glyphs, we implemented a picking mechanism that shows a tooltip with all nine values encoded in the selected glyph. We moreover implemented panning and zooming for basic interactions.

\CHANGE{We also considered including a fourth region in the glyph encoding the likelihood of a regular point (see supplement). However, as most vertices typically have a high probability of being regular, this results in a lot of visual clutter, causing visual stress. Moreover, the presence of those visually dominant glyphs in large parts of the domain results in the critical points---being the actual features of interest---to get lost in the general picture.}

\begin{figure}[t]
    \centering
    % TRIM: left, bottom, right, top
    \begin{tikzpicture}
    \node[black, inner sep=0pt] (map) {\includegraphics[trim={15px 82px 245px 80px}, clip, scale=0.093]{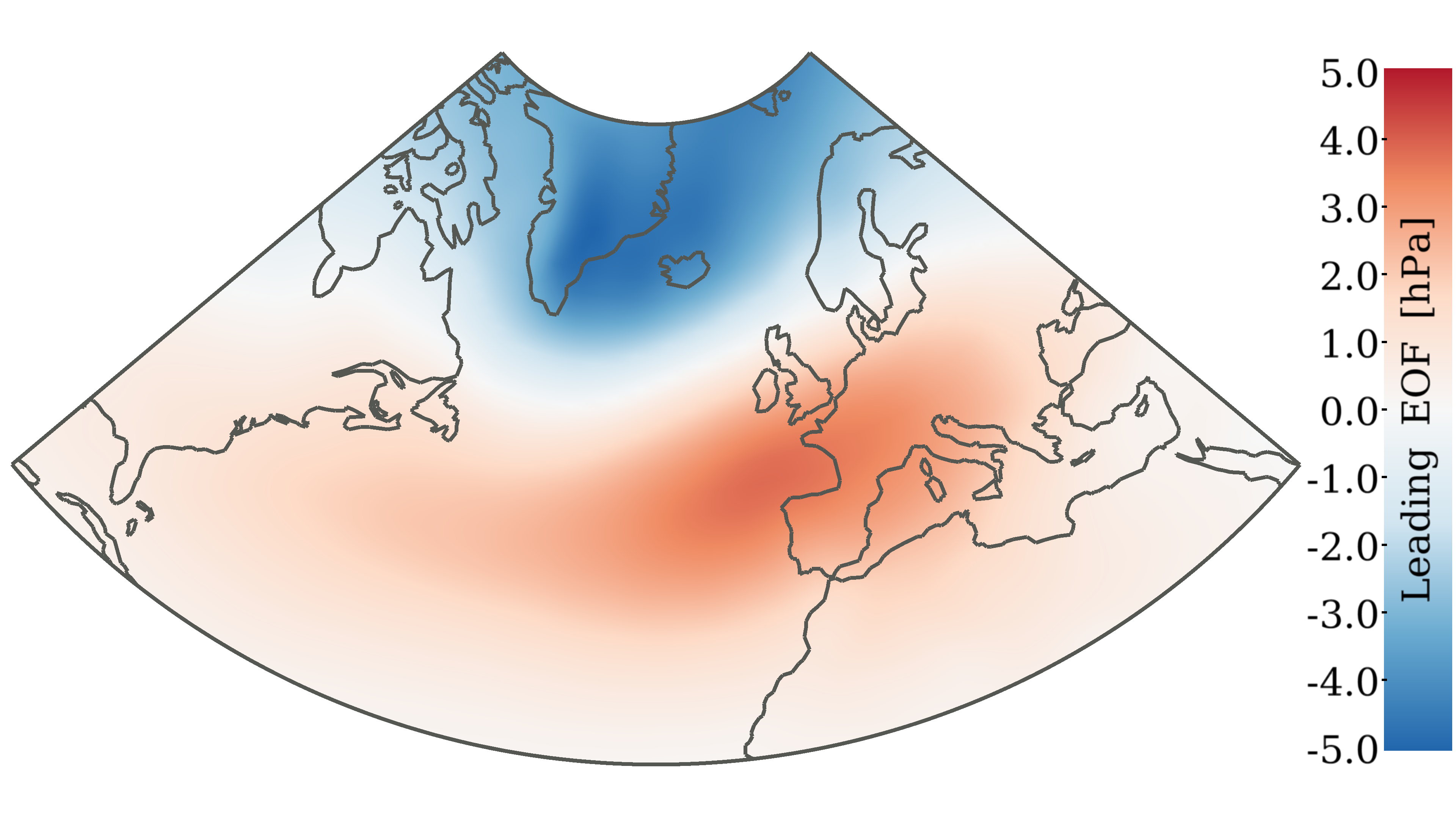}};
    \node[black, inner sep=0pt, right=1em of map] (legend) {\includegraphics[trim={2059px 82px 0px 80px}, clip, scale=0.093]{pictures/NAO_Pattern/Example_NAO_Pattern.png}};
    \end{tikzpicture}
    \caption{Leading EOF mode of sea-level pressure fields computed over the depicted region and the winters of 2029/30--2078/79 featuring the typical NAO pattern over the North Atlantic.}
    \label{fig:NAO-pattern}
\end{figure}

\section{Results}
\label{sec:results}
In the following, we validate the theoretical properties of our method on a synthetic data set and show its applicability in the context of a real-world data set from climate research.
We used a confidence level of $\gamma = 95\%$ for all presented results.

\paragraph{Synthetic Data Set}
In previous work~\cite{Vietinghoff2022}, we described a drawing process for ensemble data sets of arbitrary sizes, which we used to produce data for a comparison with existing methods for critical point detection in uncertain fields. 
Here we use the synthetic data obtained from that process to allow for a direct comparison with our own and other existing works. 
For the reader's convenience, we recapitulate here the basic idea of this drawing process:

Starting with an ensemble of annual mean sea-level pressure fields $\vec{f}_1,\dots, \vec{f}_{100}$, we estimate the mean and covariance matrix of the underlying stochastic process. 
We can then draw an arbitrary number of fields from a multivariate normal (MVN) distribution with those estimated parameters.
This drawing process has the advantage of preserving a major part of the spatial autocorrelation of the input data, providing a more representative data set than a purely synthetic one. 
For each ensemble size $m \in \{k^2 \mid k = 2,\dots,10\}$ we draw ten ensembles from the MVN distribution.

We applied the proposed method to each ensemble and produced an animation for each ensemble size to visually assess the stability of the results. 
We also produced an image for the ground truth probabilities $\ptype$ by drawing glyphs with $\hat{p}_\text{type}, \ptype^{\text{lower}}, \ptype^{\text{upper}}=\ptype$ (cf. supplement) and prepended it to each animation. 
Figure~\ref{fig:teaser} shows the pictures obtained for a single ensemble of size 9 (left) and 49 (right). The animations for those and the remaining sample sizes are provided in the supplemental material.

\begin{figure*}[htp!]
 \centering
   \begin{tikzpicture}
    % TRIM: left, bottom, right, top
      \node[draw, black, thick, inner sep=0pt] (low_begin) {%
        \includegraphics[scale=0.098]{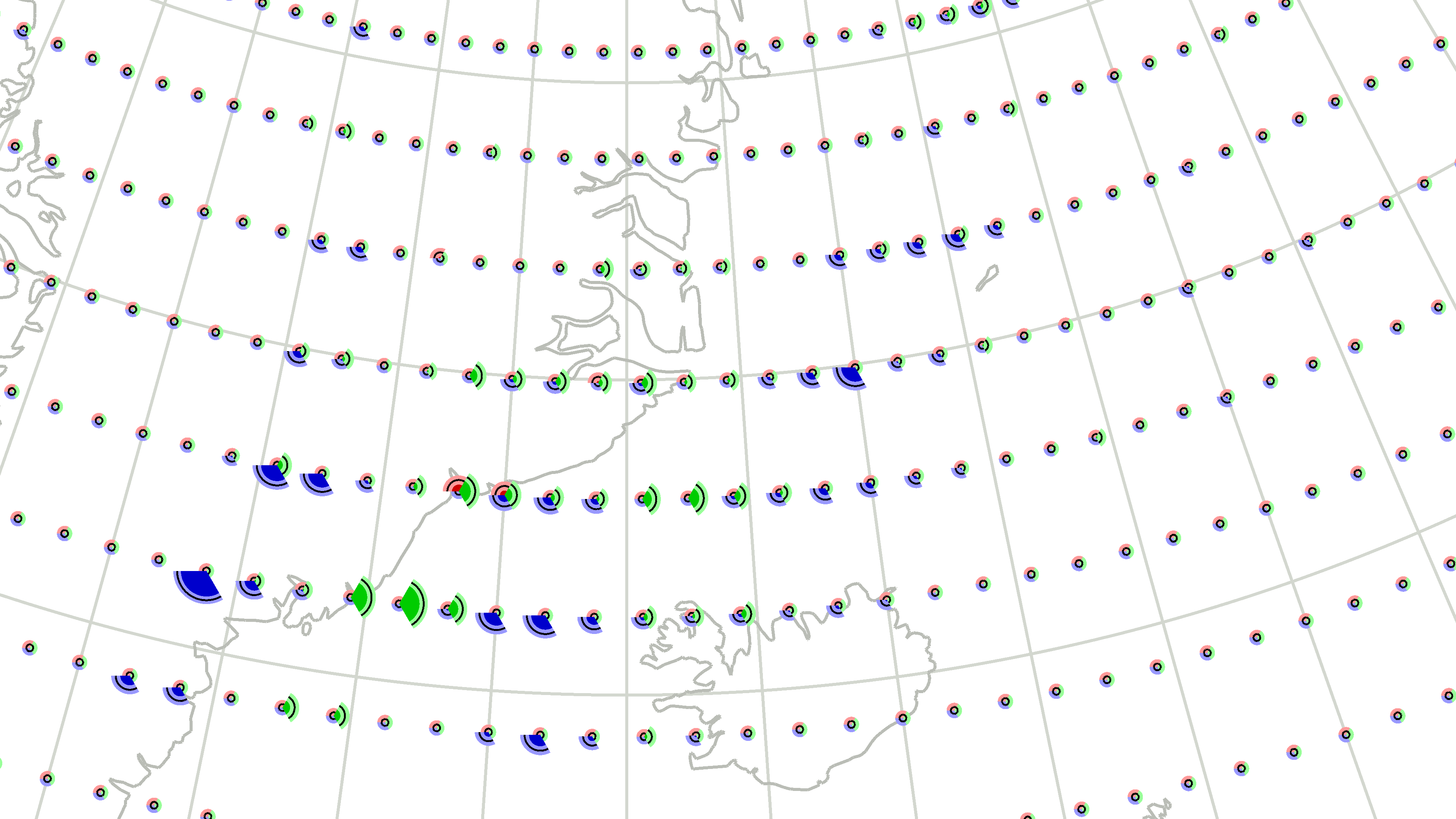}};
      \node[draw, black, thick, inner sep=0pt, right=.5em of low_begin] (low_end) {%
        \includegraphics[scale=0.098]{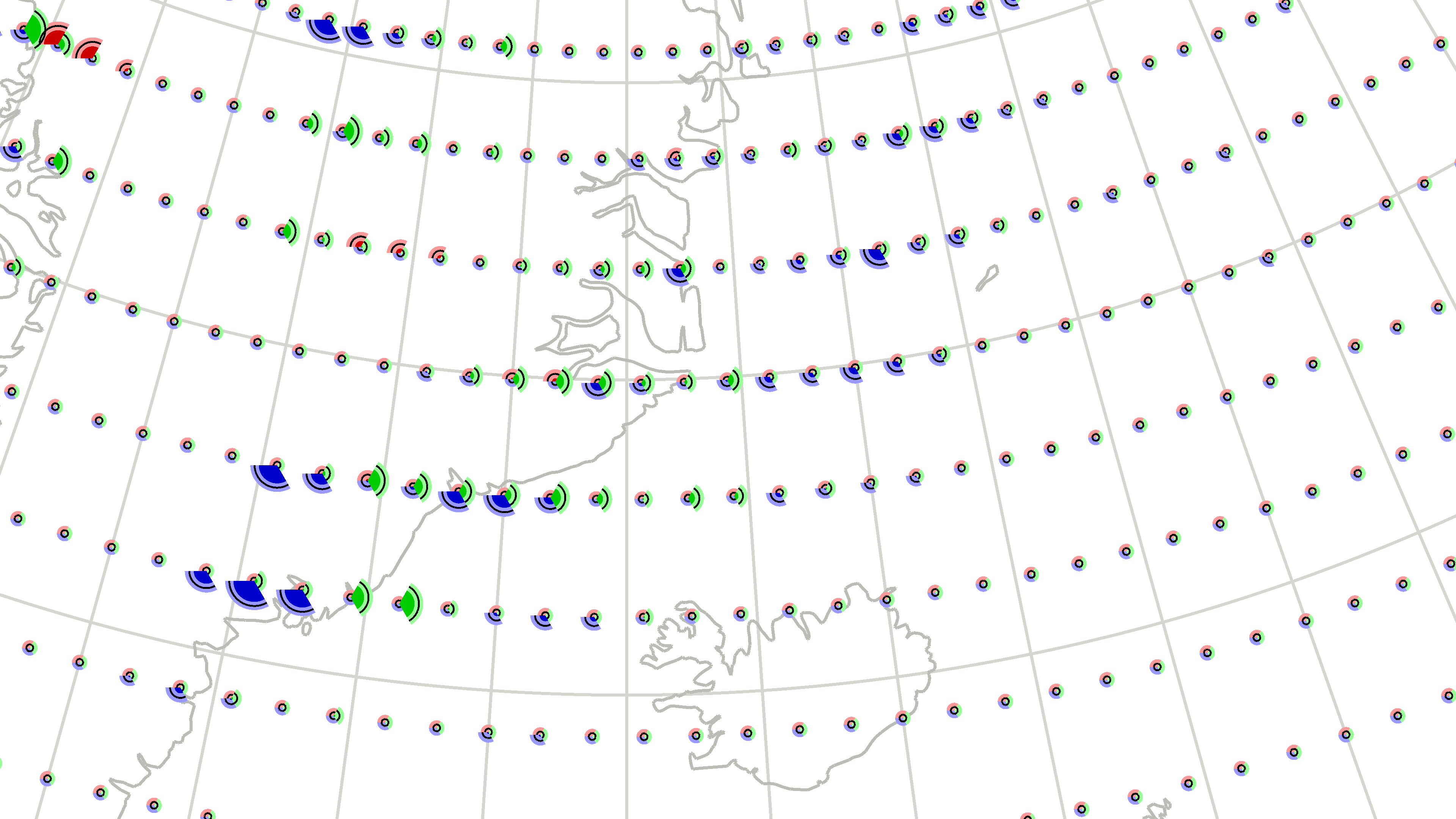}};
      \node[draw, black, thick, inner sep=0pt, below=.5em of low_begin] (high_begin) {%
        \includegraphics[scale=0.098]{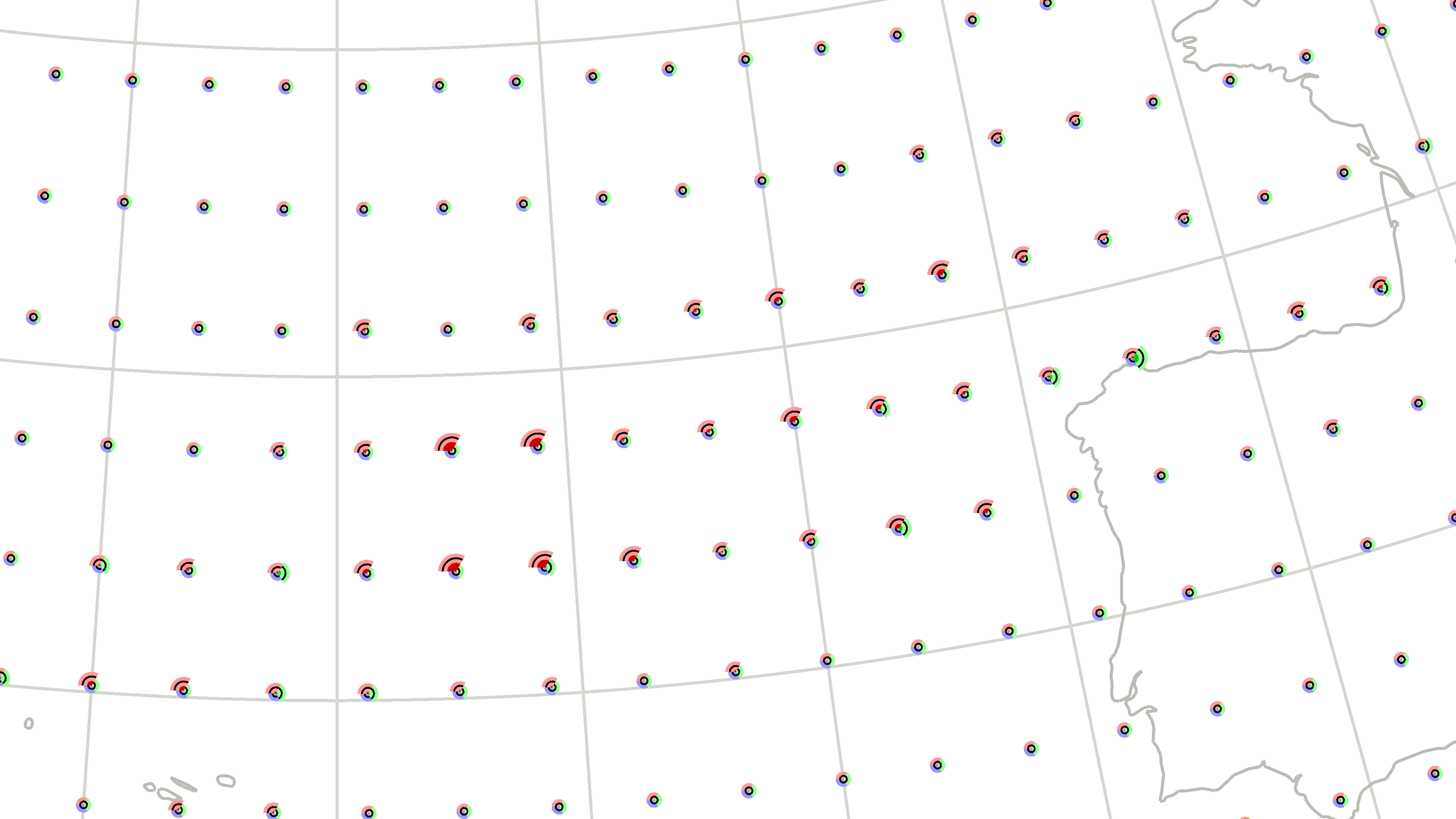}};
      \node[draw, black, thick, inner sep=0pt, below=.5em of low_end] (high_end) {%
        \includegraphics[scale=0.098]{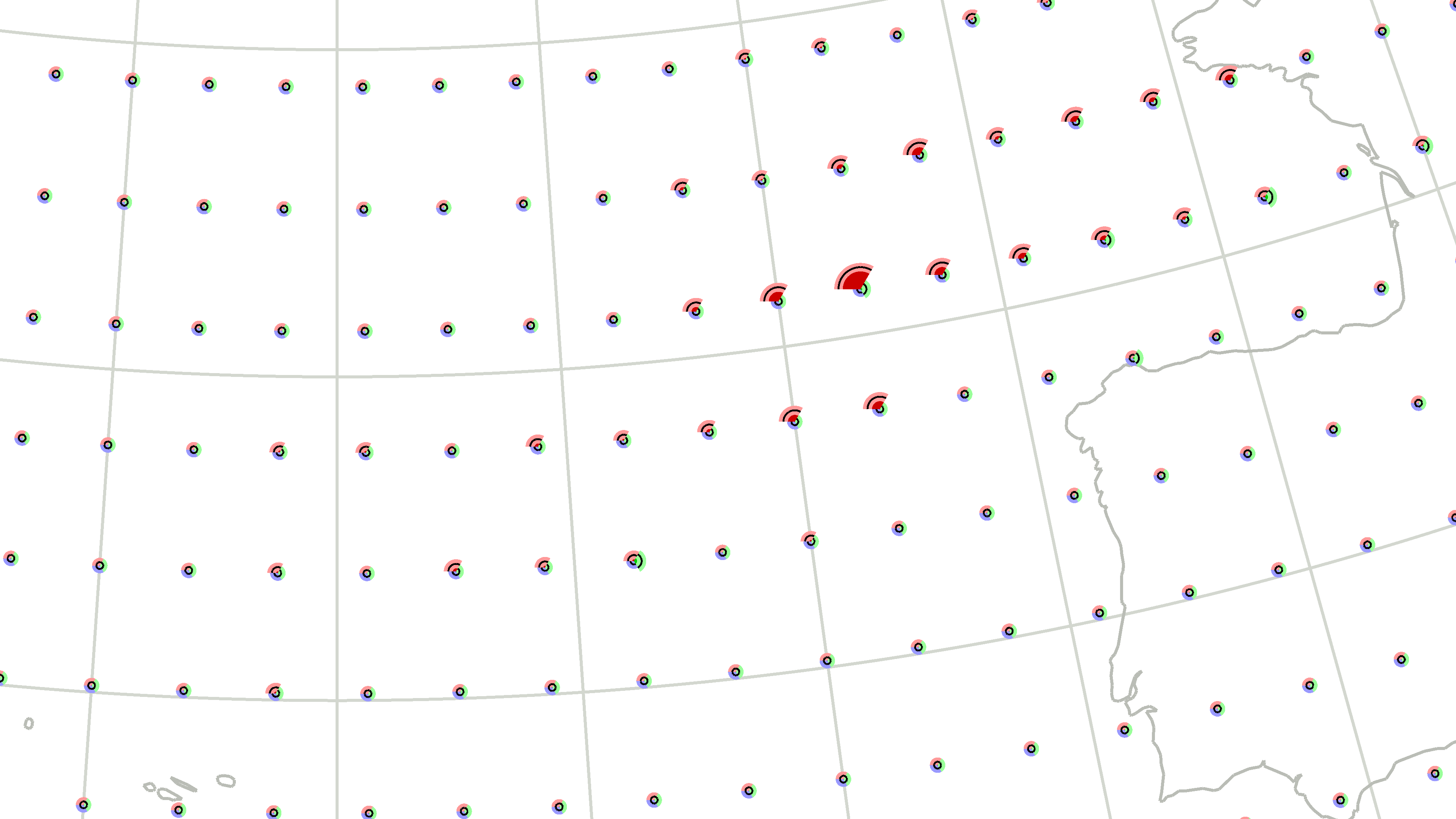}};
     
     \coordinate (vcenter) at ($0.5*(low_begin)+0.5*(high_begin)$);
     \coordinate (hcenter) at ($0.5*(low_end)+0.5*(low_begin)$);
      \node[draw, black, thick, inner sep=0pt, anchor=center] (map_overview) at (hcenter |- vcenter) {%
        \includegraphics[width=0.15\linewidth]{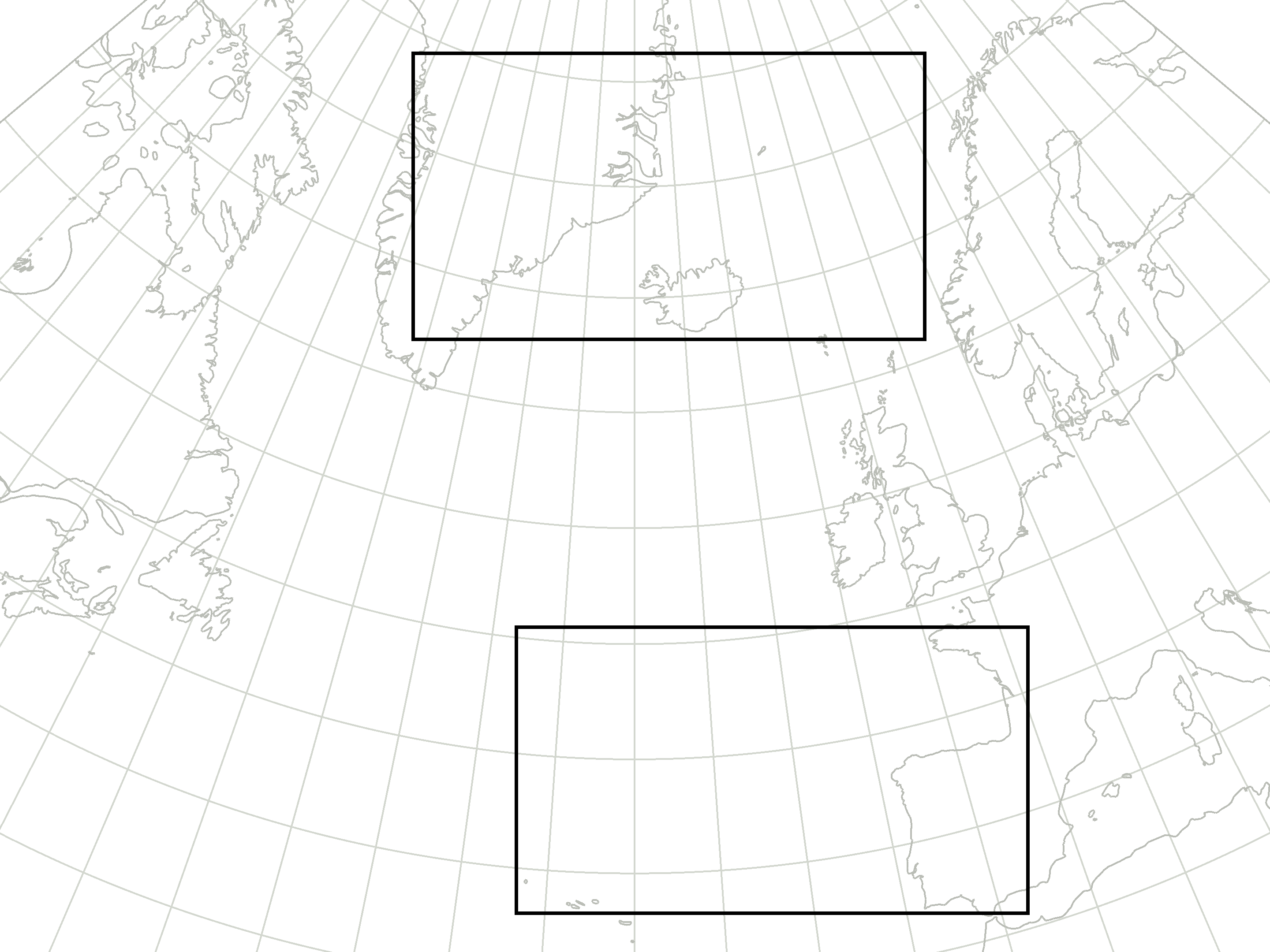}};

     \node[draw, black, thick, fill=white, rounded corners=2pt, below right=0.2cm and 0.2cm of low_begin.north west, anchor=north west] (years_begin_low) {1850/51--1899/1900};
     \node[draw, black, thick, fill=white, rounded corners=2pt, above right=0.2cm and 0.2cm of high_begin.south west, anchor=south west] (years_begin_high) {1850/51--1899/1900};  
     \node[draw, black, thick, fill=white, rounded corners=2pt, below left=0.2cm and 0.2cm of low_end.north east, anchor=north east] (years_end_low) {1949/50--1998/99};  
     \node[draw, black, thick, fill=white, rounded corners=2pt, above left=0.2cm and 0.2cm of high_end.south east, anchor=south east] (years_end_high) {1949/50--1998/99};     
     
     \node[draw, black, rounded corners=2pt, minimum width=1cm, minimum height=4.5cm, anchor=west] (legend_outline) at ($(vcenter -| low_end.east) + (.5em, 0)$) {};
     \node[below=0.3em of legend_outline.north, anchor=north, align=center, text width=1cm] (legend_header) {\textbf{Glyph Size}};
     
     \node[inner sep=0pt, below=0.5em of legend_header.south, anchor=north] (legend25) {%
        \includegraphics[angle=-90, trim={588px 302px 1634px 855px},clip,scale=0.1]{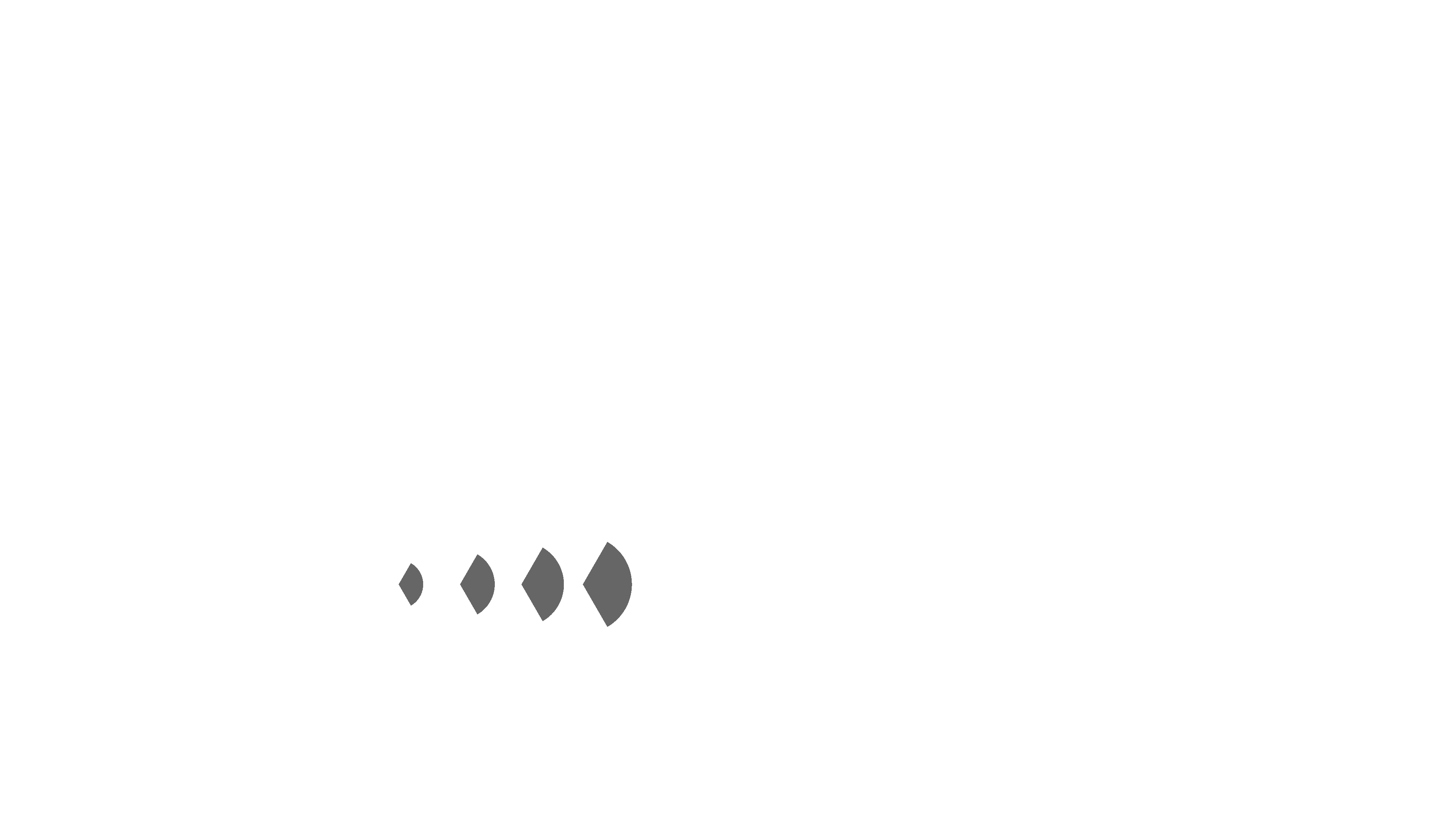}}; 
     \node[below=0em of legend25.south, anchor=north] (legend25label) {25\%};

     \node[inner sep=0pt, below=0.5em of legend25label.south, anchor=north] (legend50) {%
        \includegraphics[angle=-90,trim={702px 302px 1520px 855px},clip,scale=0.1]{pictures/Results/NAO/1pctCO2/NAO_Legend_High.png}};
     \node[below=0em of legend50.south, anchor=north] (legend50label) {50\%};
        
     \node[inner sep=0pt, below=0.5em of legend50label.south, anchor=north] (legend75) {%
        \includegraphics[angle=-90, trim={812px 302px 1410px 855px},clip,scale=0.1]{pictures/Results/NAO/1pctCO2/NAO_Legend_High.png}};
     \node[below=0em of legend75.south, anchor=north] (legend75label) {75\%};
        
     \node[inner sep=0pt, below=0.5em of legend75label.south, anchor=north] (legend100) {%
        \includegraphics[angle=-90, trim={922px 302px 1301px 855px},clip,scale=0.1]{pictures/Results/NAO/1pctCO2/NAO_Legend_High.png}};
     \node[below=0em of legend100.south, anchor=north] (legend100label) {100\%};
    \end{tikzpicture}
  \caption{Confidence intervals for critical point occurrence probabilities for an ensemble of sea-level pressure fields over the North Atlantic, indicating the northern (top row) and southern (bottom row) centers of action of the winter-time North Atlantic Oscillation at the beginning (left column) and end (right column) of a 150-years long global warming simulation with an increase in the CO$_2$ concentration of 1\% per year.
  }
  \label{fig:results-real-world}
\end{figure*}

In the result computed from only nine input fields (Figure~\ref{fig:teaser}, left), all glyphs show a comparatively large uncertainty in the form of broad outer arcs in lighter colors for all three types of critical points. In the inset, most glyphs do not contain darker-colored inner segments (e.g., \circled{\myLabelA}), meaning that the corresponding CIs start at zero. That is, it cannot be ruled out (on a 95\% confidence level) that the probability for that critical point type is zero.
For points showing glyphs with a large inner segment of darker color (e.g., \circled{\myLabelB}), on the other hand, we may be confident that they will be of the corresponding type with non-negligible probability.
More precisely, the probability of a false positive (i.e., the point in truth is not of the indicated type) is less than 2.5\%. Overall, a comparison of the prediction with the ground truth probabilities for critical points (see supplement) shows that points with non-zero lower CI boundaries (i.e., glyphs containing dark segments) had indeed a high probability of being of that type.

By increasing the number of input fields to 49 (Figure~\ref{fig:teaser}, right), the CIs shrink to much smaller ribbons. 
That is, with increasing number of samples, more accurate predictions on the critical point probabilities can be made while still maintaining the same 5\% chance of missing the true probability. A comparison with the ground truth probabilities confirms that points found to have a high probability for certain critical types were indeed likely to have that type.

What is more, the animations show that the amount of variation across the repeated draws for a single ensemble size is low for all ensemble sizes, especially in comparison with the animations obtained for the methods that do not take epistemic uncertainties into account~\cite{Mihai2014, Guenther2014, Liebmann2016} (cf. supplement of our previous work~\cite{Vietinghoff2022}).

\paragraph{Real-World Data Set}
We further applied our method to a real-world data set from climate research with the goal to investigate how climate change impacts the inter-annual variability of sea-level pressure over the North Atlantic. The most dominant mode of variation, also called the leading \emph{empirical orthogonal function} (EOF), for this region is known to characterize the North Atlantic Oscillation (NAO)~\cite{Hurrell2003}. Its characteristic spatial pattern, consisting of two anticorrelated regions, in the following called \emph{centers of action} (COAs), is depicted in Figure~\ref{fig:NAO-pattern}.

For our analysis, we used monthly mean sea-level pressure fields of the Max Planck Institute for Meteorology's Grand Ensemble (MPI-GE)~\cite{Maher2019}. MPI-GE comprises ensemble simulations for different climatic developments, where each ensemble consists of 100 realizations (members). A historical simulation covers the years 1850--2006. Each member of this simulation is directly succeeded by the corresponding members of three future scenario simulations for 2006--2099, which simulate different greenhouse gas increases. A fourth ensemble experiment simulates the Earth's climate under the influence of an increase in the CO$_2$ concentrations by 1\% per year for the historical period 1850--1999.

In a first step, we prepended each member of the three scenario simulations with the respective member of the historical run to obtain ensembles of time series covering the whole period 1850--2099. From each of these, as well as for the 1\% CO$_2$ simulation, we then computed the leading EOF with sliding time windows of 50 years, as we also did in our previous works~\cite{Vietinghoff2021, Vietinghoff2022}.
This gives us, for each time window, an ensemble of 100 EOF fields, on each of which we applied the proposed method. This results in time series of images encoding the CIs for critical point probabilities with our glyph design. 
These were then concatenated into animations to analyze how the location of the likely positions of the northern and southern COA of the NAO changes over time. 

Figure~\ref{fig:results-real-world} shows the resulting pictures for the 1\% CO$_2$ run for the first (1850/51--1899/1900, left column) and last (1949/50--1998/99, right column) time window.
From those pictures, we can identify a pronounced northeastward shift of the southern COA (bottom row) and a marginal northward trend of the northern COA (top row), both of which are in agreement with our previous findings~\cite{Vietinghoff2021, Vietinghoff2022}. Overall, the pictures for the northern COA show more glyphs with large dark portions, especially at the left-most region of minima over Greenland, indicating that with 95\% confidence the probability of minima at those points is relatively high. The southern COA, on the other hand, shows only comparatively low probabilities for maxima in a widespread region, which we also observed previously~\cite[Figure 9]{Vietinghoff2022}. 
This indicates a high amount of spatial uncertainty on the precise location of the maxima across the ensemble members. 

The animation for this and the remaining scenario simulations are provided in the supplemental material.
\section{Conclusion and Future Work}
\label{sec:conclusion}

In this work, we presented an approach for the computation and visualization of confidence intervals for the occurrence probabilities of critical points in ensembles of piecewise-linear 2D scalar fields.

We applied the method to synthetic data that has been used previously~\cite{Vietinghoff2022} to compare existing methods for the detection of critical points in uncertain data. Other than most of those methods, the approach presented here incorporates both aleatoric and epistemic uncertainties and communicates them to the viewer, thereby preventing him from trusting in predictions computed from small sample sizes that were merely a product of chance. 

We also applied our method to a data set from climate research to investigate the impact of global warming on the NAO. We could reiterate our previous findings~\cite{Vietinghoff2021, Vietinghoff2022} of systematic shifts of the two COAs defining the NAO that are more pronounced for more extreme climate change scenarios. As in our previous work~\cite{Vietinghoff2022}, we also found that the southern COA exhibits strong spatial uncertainties \CHANGE{resulting in overall low probabilities.
This effect gets worse with increasing grid resolutions because the perturbed nature of the members will then make it more likely that the critical points are located at slightly different grid points.
This is a limitation of methods that analyze the criticality of a field on a per-vertex basis.
We hence plan to search for approaches that also consider the spatial autocorrelation in the data to identify regions of critical points with spatial uncertainty as the same features. This would mitigate the dependency of the results from the grid resolution as the accumulated probability of each feature would not be as strongly affected by the resolution. 
Another point to keep in mind is that the existence and position of topological features of PL fields depends on the way the domain is triangulated~\cite{Carr2006, Scheuermann1998}. Finally, as noise in (real-world) data sets can lead to a large number of critical points with low persistence~\cite{Tierny2017}, a preprocessing step~(e.g., \cite{Tierny2012}) might be required to avoid artifacts.}

CIs are a common means for capturing the (epistemic) uncertainties that arise from small sample approximations. However, studies have shown that CIs can be prone to misinterpretations~\cite{Hoekstra2014}. An evaluation thus seems advisable to assess how well the here proposed glyphs are interpreted by users with different backgrounds. Overall, it might be interesting to examine how CIs can be visually represented to mitigate such common misinterpretations.

\acknowledgments{The authors wish to thank the Deutsche Forschungsgemeinschaft for funding the project SCHE 663/11-2.}
\newpage

\bibliographystyle{abbrv-doi-hyperref}

\bibliography{template}
\end{document}